\newcommand{\lya}{\mbox{Ly$\alpha$}}

\newcommand{\halpha}{\mbox{H$\alpha$}}
\newcommand{\hbeta}{\mbox{H$\beta$}}

\newcommand{\cIV}{\mbox{C~{\sc iv}}}

\newcommand{\siIV}{\mbox{Si~{\sc iv}}}

\newcommand{\nV}{\mbox{N~{\sc v}}}

\newcommand{\kms}{\mbox{km\,s$^{-1}$}}

\newcommand{\msun}{\mbox{M$_\odot$}}

\newcommand{\fcov}{\mbox{$f_\mathrm{cov}$}}

\newcommand{\mstell}{\mbox{$M_\mathrm{stell}$}}

\documentclass[fleqn,usenatbib]{mnras}

\usepackage{newtxtext,newtxmath}
\usepackage{hyperref}
\usepackage{xcolor}

\usepackage[T1]{fontenc}

\DeclareRobustCommand{\VAN}[3]{#2}
\let\VANthebibliography\thebibliography
\def\thebibliography{\DeclareRobustCommand{\VAN}[3]{##3}\VANthebibliography}

\usepackage{graphicx}	
\usepackage{amsmath}	

\title[Accelerating starburst winds]{Accelerating Galaxy Winds During the Big Bang of Starbursts}

\author[M. J. Hayes]{Matthew. J. Hayes$^{1}$\thanks{E-mail: matthew@astro.su.se (MJH)} \\
$^{1}$Stockholm University, Department of Astronomy and Oskar Klein Centre for Cosmoparticle Physics, AlbaNova University Centre, SE-10691, Stockholm, Sweden.}

\date{Accepted XXX. Received YYY; in original form ZZZ}

\pubyear{2022}

\begin{document}
\label{firstpage}
\pagerange{\pageref{firstpage}--\pageref{lastpage}}
\maketitle

\begin{abstract}
We develop a new method to infer the temporal, geometric, and energetic properties of galaxy outflows, by combining stellar spectral modeling to infer starburst ages, and absorption lines to measure velocities.  If winds are accelerated with time during a starburst event, then these two measurements enable us to solve for the wind radius, similarly to length scales and the Hubble parameter in Big Bang cosmology. This wind radius is the vital, but hard-to-constrain parameter in wind physics. We demonstrate the method using spectra of 87 starburst galaxies at $z=0.05-0.44$, finding that winds accelerate throughout the starburst phase and grow to typical radii of $\approx 1$~kpc in $\approx 10$~Myr.  Mass flow rates increase rapidly with time, and the mass-loading factor exceeds unity at about 10~Myr -- while still being accelerated, the gas will likely unbind from the local potential and enrich the circumgalactic medium.  We model the mechanical energy available from stellar winds and supernovae, and estimate that a negligible amount is accounted for in the cool outflow at early times. However, the energy deposition increases rapidly and $\sim 10$\% of the budget is accounted for in the cool flow at 10~Myr, similar to some recent hydrodynamical simulations.  We discuss how this model can be developed, especially for high-redshift galaxies. 
\end{abstract}


\section{Introduction}\label{sect:intro}

It is impossible to overstate the role played by feedback and large-scale winds in galaxy evolution.  Winds drive material out from galaxies, regulating the fuel available for star formation. In the same process winds redistribute metals, preferentially removing them from the ISM to circumgalactic and possibly intergalactic media.  Many simulations address galaxy winds, and numerous feedback mechanisms can be simulated at scales that well-resolve molecular clouds. However as larger volumes are tackled, more of the physics is re-implemented, and approximated in a computationally tractable `sub-grid' manner (see \citealt{Somerville2015} for a review). For simulations that address large galaxy populations over cosmological volumes, most of the feedback physics is implemented in this way.

Empirical calibration of these recipes is needed, specifically addressing (for example) how the star formation process truly maps onto gas heating and acceleration.  This is especially true when considering energy transfer over galaxy scales and how mass, metals, and angular moment are redistributed.  A common observational approach, especially for ultraviolet-bright star-forming galaxies, is to perform absorption line spectroscopy along sight-lines to the brightest UV sources. At low redshifts, such studies have demonstrated the empirical scaling of wind properties (e.g. outflow velocities, mass-loading factors; $\eta\equiv \dot M_\mathrm{out} / \mathrm{SFR}$) with physical galaxy properties of SFR and stellar mass 
\citep[e.g.][]{Heckman.2015,Chisholm.2017,Xu.2022classy}. 

At some step in any such calculation, all line-of-sight strategies must convert an estimated column density into a number of particles, and therefore a mass.  Momentum and energy flow rates then follow directly.  This requires an estimate of the surface area over which the wind is operating, which in practice means we need a radius. In the spherically symmetric case with a thin shell, this would be $M_\mathrm{wind} = 4\pi r_\mathrm{wind}^2 f_\mathrm{cov} N_\mathrm{H} m_\mathrm{H} \mu$, where $r_\mathrm{wind}$ is the radius of the wind, \fcov\ is the covering fraction, $N_\mathrm{H}$ is the column density of hydrogen atoms, $m_\mathrm{H}$ is the mass of the hydrogen atom, and $\mu$ is the mean molecular weight.  

Knowing the position of material along the line of sight ($r_\mathrm{wind}$, here) is a well-known difficulty in absorption line spectroscopy, as location is not directly encoded.  Various solutions to this have been presented, such as linking wind size to that of the starburst region \citep[e.g.][]{Heckman.2015}, deriving a distance from photoionization modeling \citep[e.g.][]{Chisholm.2017}, or expressing $r$ in continuity equations to describe an envelope \citep[e.g.][]{Carr.2018}.  In this \emph{Letter} we outline a new technique by which the radial distance between the light source and the wind can be estimated.  The method relies upon representative samples of galaxies being observed at different evolutionary times throughout the starburst. We describe the sample in Section~\ref{sect:sample} and the method in Section~\ref{sect:methods}.  We present the main results -- derivations of the mass flow rates, mass loading factor, and wind energy -- in Section~\ref{sect:results}.  In Section~\ref{sect:disc} we present a brief discussion of how the method can be improved and developed at high-redshifts.  

\begin{figure}
	\includegraphics[width=0.49\columnwidth]{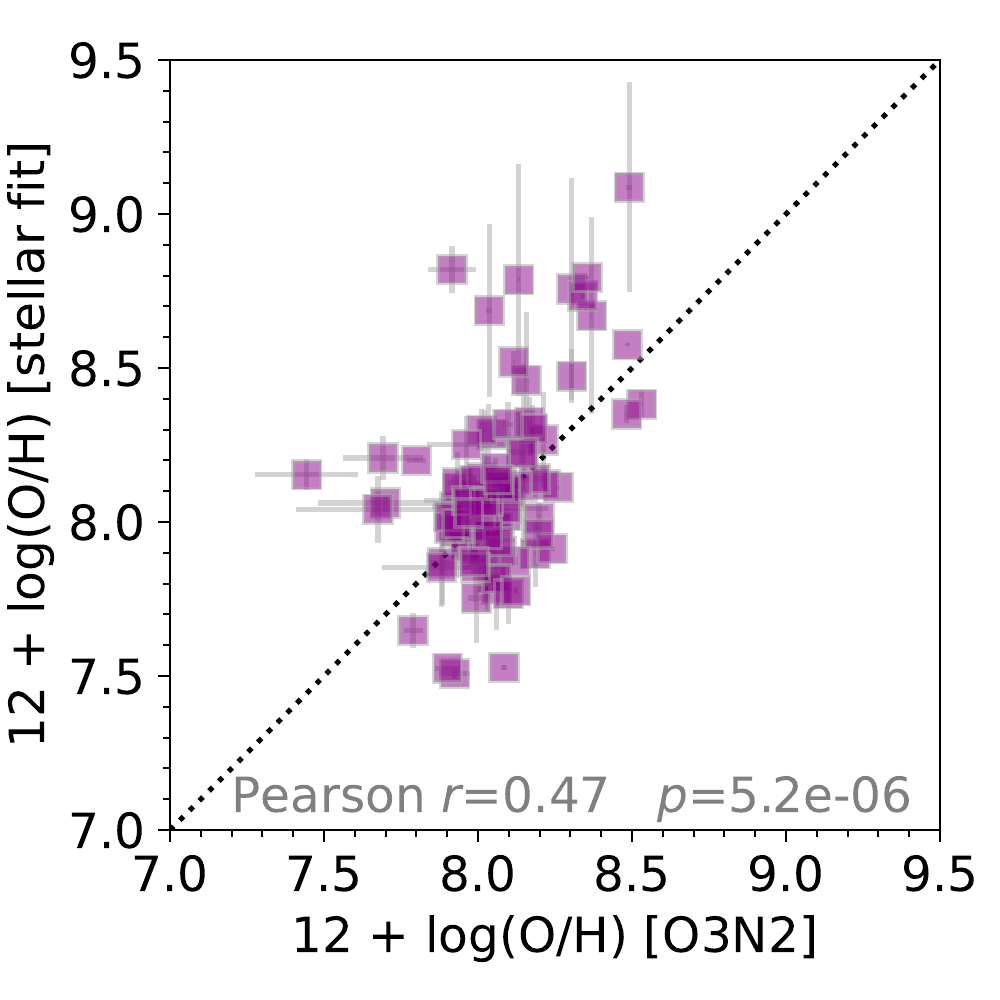}
	\includegraphics[width=0.49\columnwidth]{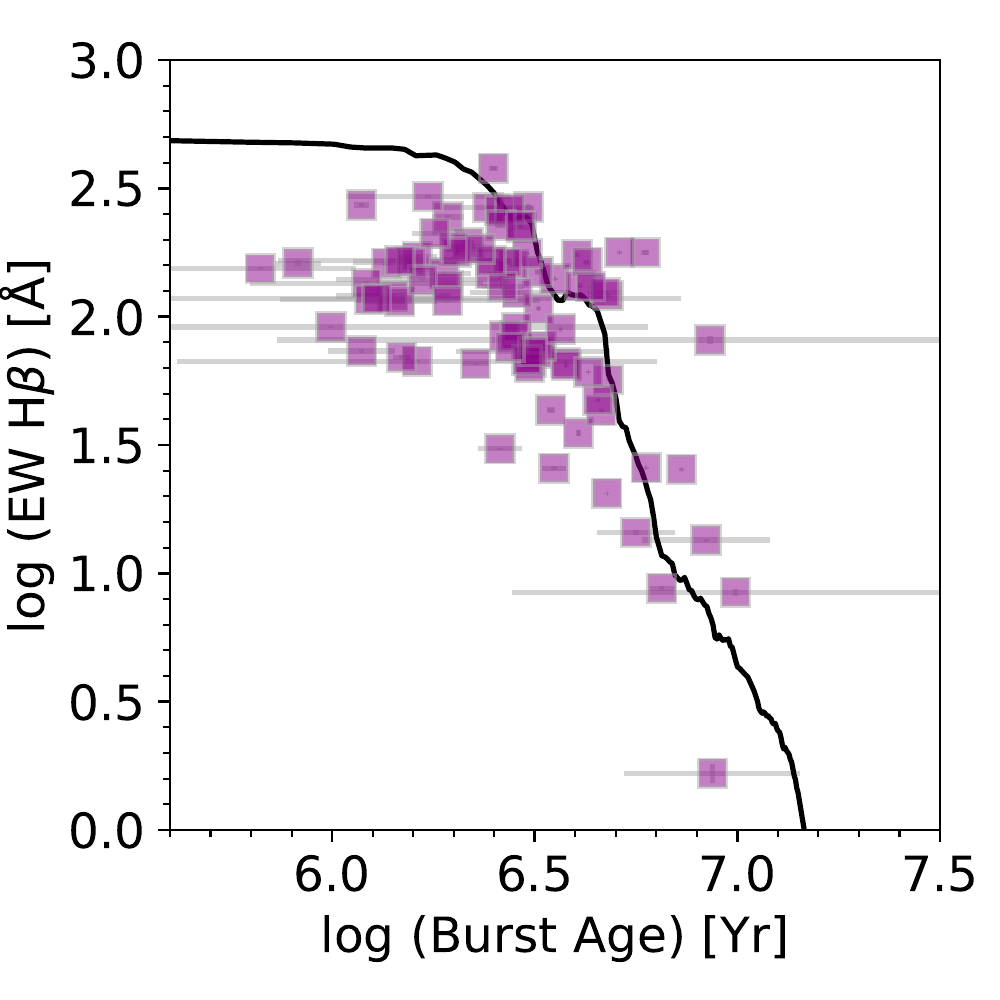}
    \caption{General results of the spectral modeling.  Left shows the stellar metallicity recovered from fitting \citep[converted to oxygen abundance using][]{Asplund.2009}, compared to the gas-phase abundance derived from ([\ion{O}{iii}]5007/\hbeta)/([\ion{N}{ii}]6584/\halpha) following  \citet{Marino.2013}.  Right: the \hbeta\ equivalent width shown against the log of the burst age recovered from the fitting.  The solid line shows the expected EW based upon recombination theory and no differential reddening \citep{Leitherer.1999}. }
    \label{fig:modelprop}
\end{figure}

\section{Sample}\label{sect:sample}

The sample is drawn from a full archive of observations of starburst galaxies observed with the Cosmic Origins Spectrograph \citep[COS;][]{Green.2012} onboard the Hubble Space Telescope (HST).  Almost 200 galaxies have been observed since COS was mounted, yielding a diverse sample that has many applications.  We maintain a large database of these spectra, where we include only objects that have been observed with \ion{H}{i} \lya\ in the G130M or G160M gratings.  The samples have been selected with numerous different requirements, such as studying the UV continuum and absorption lines directly \citep[e.g.][]{Heckman.2011}, the \lya\ emission line \citep[e.g.][]{Henry.2015}, or the emission of ionizing radiation \citep[e.g.][]{Izotov.2018}. All these campaigns rejected Active Galactic Nuclei using at least optical line widths and flux ratios. The sample is generally recognized as being comparable to Lyman break galaxies and strong emission line galaxies reminiscent of objects at early times.  A subset of the sample was used in \citet{Hayes.2021} and the full properties will be presented in a forthcoming paper (Hayes et al., 2022 in preparation).  

The redshift range of the parent sample is $z\leq 0.44$.  Here we impose an additional cut of $z>0.05$ in order to capture a large fraction of the stellar light with COS: this reduces the total sample to 87 galaxies.  Even the most massive galaxies in the sample have stellar masses of the order $10^9$~\msun, and are significantly less massive than the Milky Way.  We examine their sizes in the $u-$band using the \emph{PetroR50} radius in the \texttt{PhotoObj} table of the Sloan Digital Sky Survey databases \citep{SDSS.DR17}: the distribution of \emph{PetroR50} peaks at  0\farcs7, and all but three galaxies have half-light radii smaller than the 1\farcs25 aperture of the COS.  Moreover, the ultraviolet light measured at Hubble resolution is typically far more concentrated than that measured from the ground, and we feel confident that the great majority of the UV light is captured by the spectrograph.

\section{Description of the methods}\label{sect:methods}

Velocities are easy to measure from absorption spectroscopy, when there is enough cool material along the line-of-sight.  However the distance between the emitting sources and absorbing clouds cannot be obtained directly.  In this study we rely upon the argument that winds must be launched: either cool material is accelerated by pressure from hot gas or radiation, or can condense out of the hot fluid \citep[e.g.][]{Gronke.2018}.  Depending upon the application, however, the exact mechanism may be of varied importance, as in both instances the gas is static ($v_\mathrm{wind}=0$) at $t=0$ and accelerated over time.  If we knew $v_\mathrm{wind}$ at time $t$ since the onset of the starburst event, we could integrate with respect to time to obtain the radius, such that $r_\mathrm{wind}(t) = \int_0^{t_\mathrm{obs}} v_\mathrm{wind}(t) dt$. 

\subsection{Stellar modeling and basic properties of the sample}\label{sect:model}

The challenge is to estimate the evolutionary stage of the starburst episode, which we do by modeling the stellar spectra with evolutionary synthesis models.  This procedure, as well as a complete description of the sample, will be presented in Hayes et al (2022, in prep), but the engine has already been used by \citet[][see also \citealt{Chisholm.2019} for comparable methods]{Sirressi.2022}.  We adopt Starburst99 \citep{Leitherer.1999,Leitherer.2014} high resolution ultraviolet and optical models, built with Geneva tracks, and \citet{Salpeter.1955} IMF to $M_\mathrm{up}=100$~\msun. All the COS spectra have been obtained co-spatially with optical spectra from the SDSS, and we fit the full spectral range: restframe wavelengths between $\lambda=1220$~\AA\ and $\sim 8000$~\AA, with a gap around $\sim 1400-3500$\AA. Most importantly we  always sample at least one of the P\,Cygni lines that form in the atmospheres of massive stars (\nV$\lambda 1240$\AA, and sometimes \siIV$\lambda 1400$ and \cIV$\lambda 1550$\AA), the complete UV-to-optical colors, the 4000\AA\ break, and redder wavelengths where the light may be dominated by older, cooler stars.   We fit three independent stellar populations (ages $10^6$ to $10^{10}$ yr), masked for interstellar features, and reddened using the \citet{Calzetti.2000} attenuation curve.  

For every galaxy we recover the age and metallicity of each stellar population.  The degeneracy between these two properties is well-known in stellar modeling as individual absorption features reach similar amplitudes at different times as a function of metal abundance \citep[e.g.][]{Worthey.1994}.  In Figure~\ref{fig:modelprop} we show there is good agreement between the stellar and nebular abundances, suggesting that we are able to break this degeneracy reasonably well.  

We mainly seek to age-date the starburst event, but this starburst is likely to be hosted in a more evolved galaxy.  We therefore compute a `burst age', as the mass-weighted average of stellar generations with ages below 50~Myr.  The strong nebular emission lines, and the general absence of broad absorption wings in the Balmer series, suggest that the burst events in our galaxies will not have characteristic ages older than this.  This method finds two young populations in almost all cases: one is invariably drawn to ages 2--6~Myr to reproduce the stellar wind features, while a second is more flexible and captures a more evolved starburst.  The mass-weighted burst ages range from $\approx 2$ to 20~Myr.  The right panel shows the relation between burst age and the \hbeta\ equivalent width (EW), where the anti-correlation is expected.  The corresponding theoretical line from Starburst99 is overlaid, calculated for an instantaneous burst with metallicity $Z=0.008$.  The data approximately follow the shape of the theoretical curve, but generally fall below it.  The lower observed EWs are a natural consequence of the a more evolved population contributing in the optical continuum, and differing extinctions between nebular light and starlight \citep[e.g.][]{Calzetti.2000}.  We believe that the modeled evolutionary timescales are reliable enough to usefully characterize `starburst ages'.

\begin{figure}
	\includegraphics[width=0.99\columnwidth]{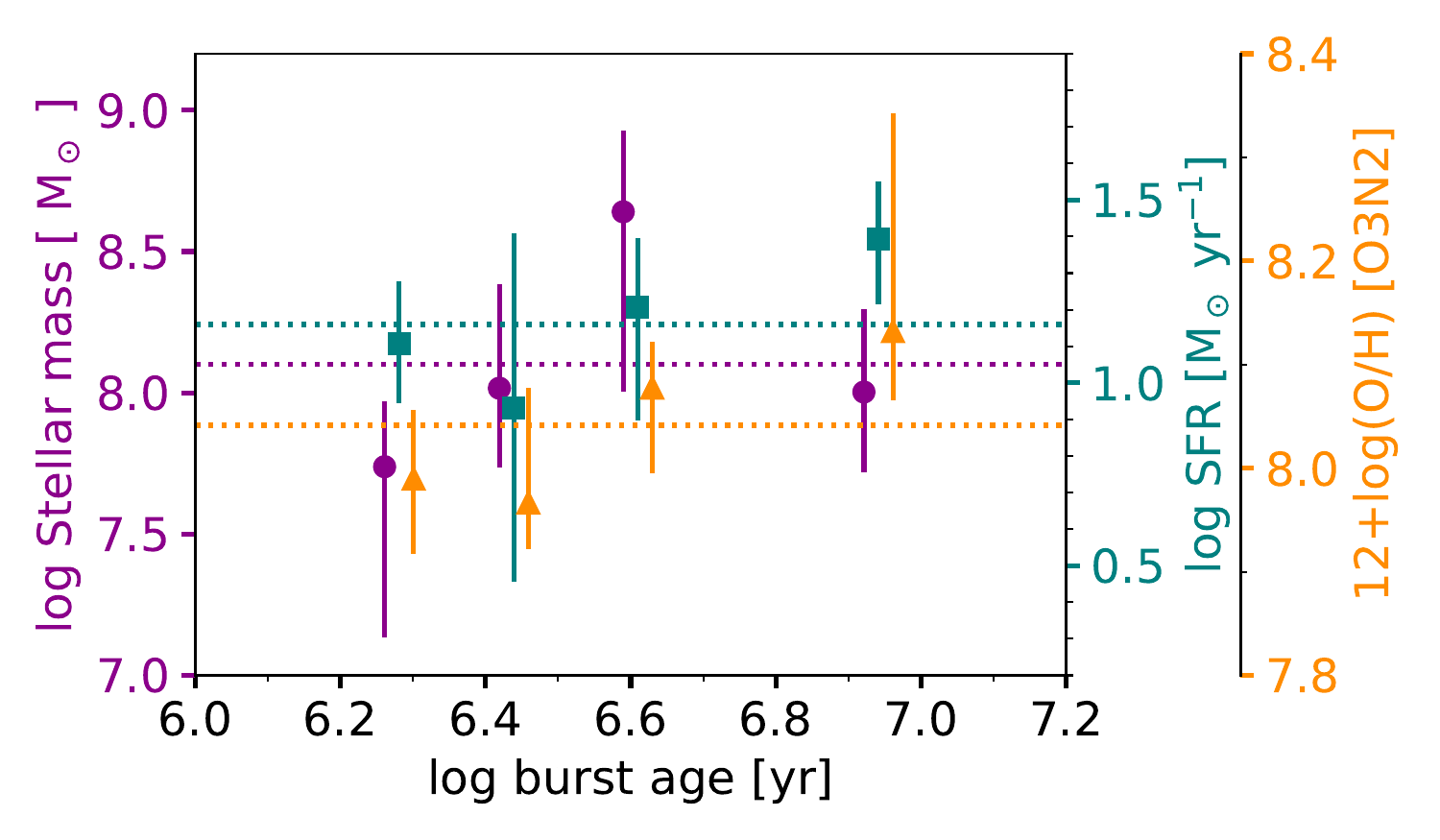}
    \caption{Average physical properties of the galaxies.  The sample is binned into four equally sized bins by starburst age, but are slightly offset in the x-direction for clarity.  Different colors show stellar mass (magenta), SFR estimated from dust-corrected \halpha\ \citep[teal;][]{Kennicutt.2012}, and nebular metallicity estimated from the O3N2 method \citep[orange;][]{Marino.2013}.  Datapoints show the median value and errorbars show the 25--75 percentile range. Horizontal dotted lines show the median values of the full sample.}
    \label{fig:genprops}
\end{figure}

\begin{figure*}
	\includegraphics[width=0.7\columnwidth]{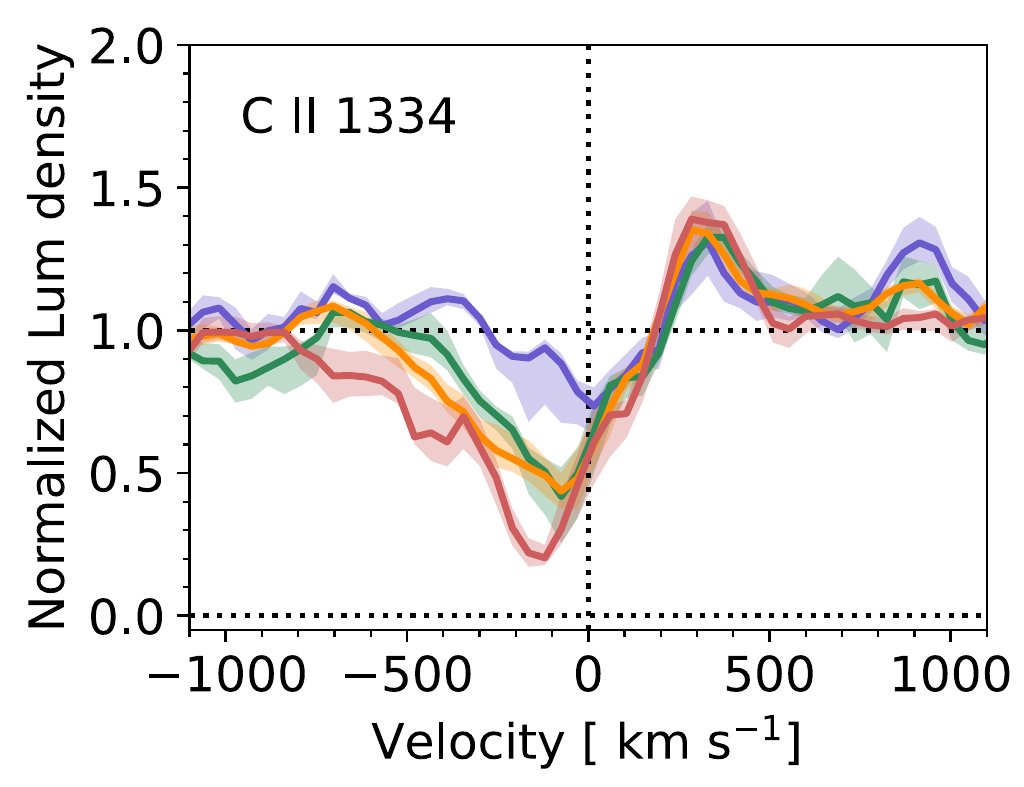}
	\includegraphics[width=1.\columnwidth]{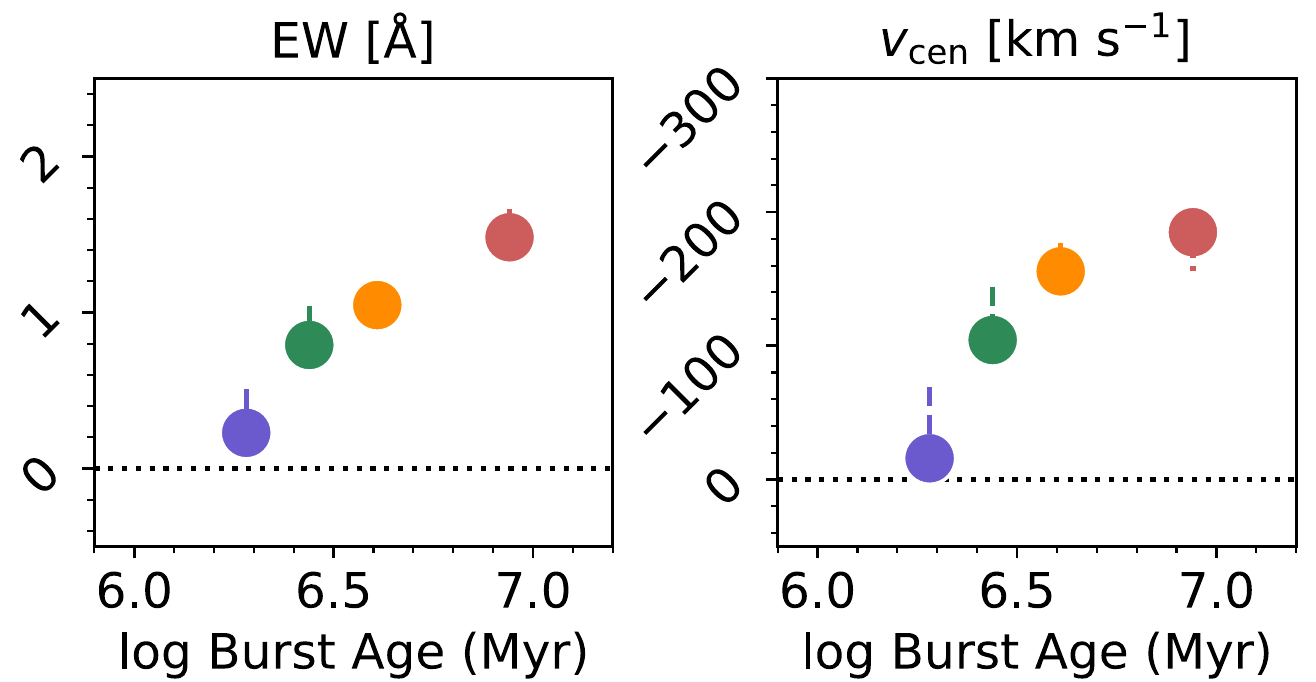}
    \caption{\emph{Left} shows the normalized and stacked UV spectra, zoomed in around the \ion{C}{ii}$\lambda 1334$\AA\ line. Bins are the same as in Fig~\ref{fig:genprops}, and show increasing starburst age, through the sequence blue, green, orange, red.  Shaded regions show the 25--75 percentiles and are derived by bootstrap analysis.  \emph{Right} panels show the median absorption equivalent width and the first moment of velocity, derived from the average spectra in the left panel.}
    \label{fig:abslines}
\end{figure*}

When we refer to stellar masses we sum the total mass of all populations recovered by the fitting algorithm, including those older than 50~Myr.  We show the average total masses (\mstell), binned by burst age in Figure~\ref{fig:genprops} (see also Section~\ref{sect:absline}), along with commonly derived properties of SFR, and nebular oxygen abundance.  Since we are concerned with young starburst events, we compute the SFR from the (dust-corrected) \halpha\ luminosity \citep{Kennicutt.2012}.  Metallicities are derived from the O3N2 index [$\equiv$~([\ion{O}{iii}]5007/\hbeta)/([\ion{N}{ii}]6584/\halpha)] following \citet{Marino.2013}.  Figure~\ref{fig:genprops} shows a hint of a trend between burst age and  \mstell, SFR, and 12+log(O/H), but the differences are less than 0.5 dex and thus we believe that secondary correlations between outflow velocity and SFR or mass will not strongly impact our analysis.

\subsection{Absorption line measurements}\label{sect:absline}

There is a natural trade-off in this study.  We demand a large sample of galaxies with UV spectra, but we also need high signal/noise ratios in the continuum to measure the absorption lines.  This is possible only for the minority of our sample, and we therefore adopt a stacking analysis to more accurately measure the individual features. 

We normalize each of the UV spectra by its best-fitting synthetic model (Section~\ref{sect:model}), sort them into four equally-sized bins as a function of starburst age, and stack the spectra in each bin. Figure~\ref{fig:abslines} shows an example of the \ion{C}{ii} line at 1334\AA.  We use the \ion{C}{ii} line here because it is one of the strongest lines visible in the wavelength coverage of the spectra, not contaminated by airglow emission like the \ion{O}{i}~$\lambda1302$\AA\ line, and less dust-depleted than silicon. Very similar results, however, are obtained when substituting the \ion{Si}{ii}~$\lambda1260$\AA\ line. Two things are immediately clear from the left-hand panel: as the inferred burst age increases, the absorption lines become both more blueshifted (based upon velocity centroid) and stronger (both in depth and EW).  These two quantities of absorption equivalent width and velocity centroid are shown as a function median burst age in the right two panels.  

\subsection{Wind model}\label{sect:windmodel}

Figure~\ref{fig:abslines} shows that at $t\approx 1$~Myr, the first moment is almost consistent with zero, and no outflow is detected.  At $t\approx 10$~Myr, however, the first moment has shifted to $\approx 200$~\kms.  It is not necessarily a surprise that the first moment shifts to more negative velocities with time: all winds must accelerate over some timeframe and the cool gas may be either driven out by the collective action of stellar winds and supernova feedback, or may condense from a fast-moving hotter fluid.  This simple result shows the relation between these feedback sources and the acceleration of gas.

If taken at face value the stacked spectra give us the wind speed as a function of time.  Assuming continuity between the subsamples, and that the initial wind radius is small (analogous to the Universe in the Big Bang model) we can integrate velocity with respect to time and obtain the radius of the wind ($r_\mathrm{wind}$).  This length is critical because it overcomes the main shortcoming of absorption line spectroscopy: the distance between light source and wind can now be constrained (Section~\ref{sect:intro}).  

We estimate the column density of gas from the EW of \ion{C}{ii} at $\lambda= 1334$\AA.  Since this line is usually saturated we develop a basic empirical calibration, using the COS observations of \citet{James.2014}.  This sample is smaller than ours (12 independent measurements) but the galaxies are much closer.  The UV continuum data are much brighter and the unsaturated triplet of \ion{S}{ii} at $\lambda=1251-1259$~\AA\ is measured.  We use the \ion{S}{ii} column densities derived in \citet{James.2014} assuming \fcov=1, and measure the \ion{C}{ii} EWs in the same way as for our main study.  A logarithmic regression gives $\log (N_\mathrm{S~II}/\mathrm{cm^{-2}}) = 0.48 W_\mathrm{C~II} + 14.73$, with an rms of only 0.14~dex. We use this relation to estimate the \ion{S}{ii} column density.  

Using a wind surface area of $4\pi r_\mathrm{wind}^2$, and assuming the outflow is of equivalent metallicity to the nebular gas (see Figure~\ref{fig:genprops}) we derive the total mass in hydrogen and helium.  We have also computed the gas column density using the extinction-based method of \citet[][see also \citealt{Heckman.2011}]{Calabro.2022} and found the two quantities agree to within 60~\%, except in the very youngest bin where the discrepancy is somewhat larger.  We note also that this calculation reduces the wind structure to one where all the absorbing gas is found at radius $r_\mathrm{wind}$, whereas in reality winds probably exhibit a more extended, clumpy, multi-phase structure, where cool material may be entrained and mixed \citep[c.f.][]{Schneider.2020}.  Further discussion of this caveat can be found in Section~\ref{sect:disc}. With the mass and velocity we proceed to compute the wind kinetic energy, which we also contrast with the mechanical energy budget estimated from spectral modeling.  With the mass outflow rate we can estimate the mass loading factor, which is important in understanding the feedback efficiency.  The focus of this Letter is to present a model by which these quantities can be derived, which we discuss in Section~\ref{sect:results}.

\section{Results -- the development of galaxy winds}\label{sect:results}

\begin{figure*}
	\includegraphics[width=2\columnwidth]{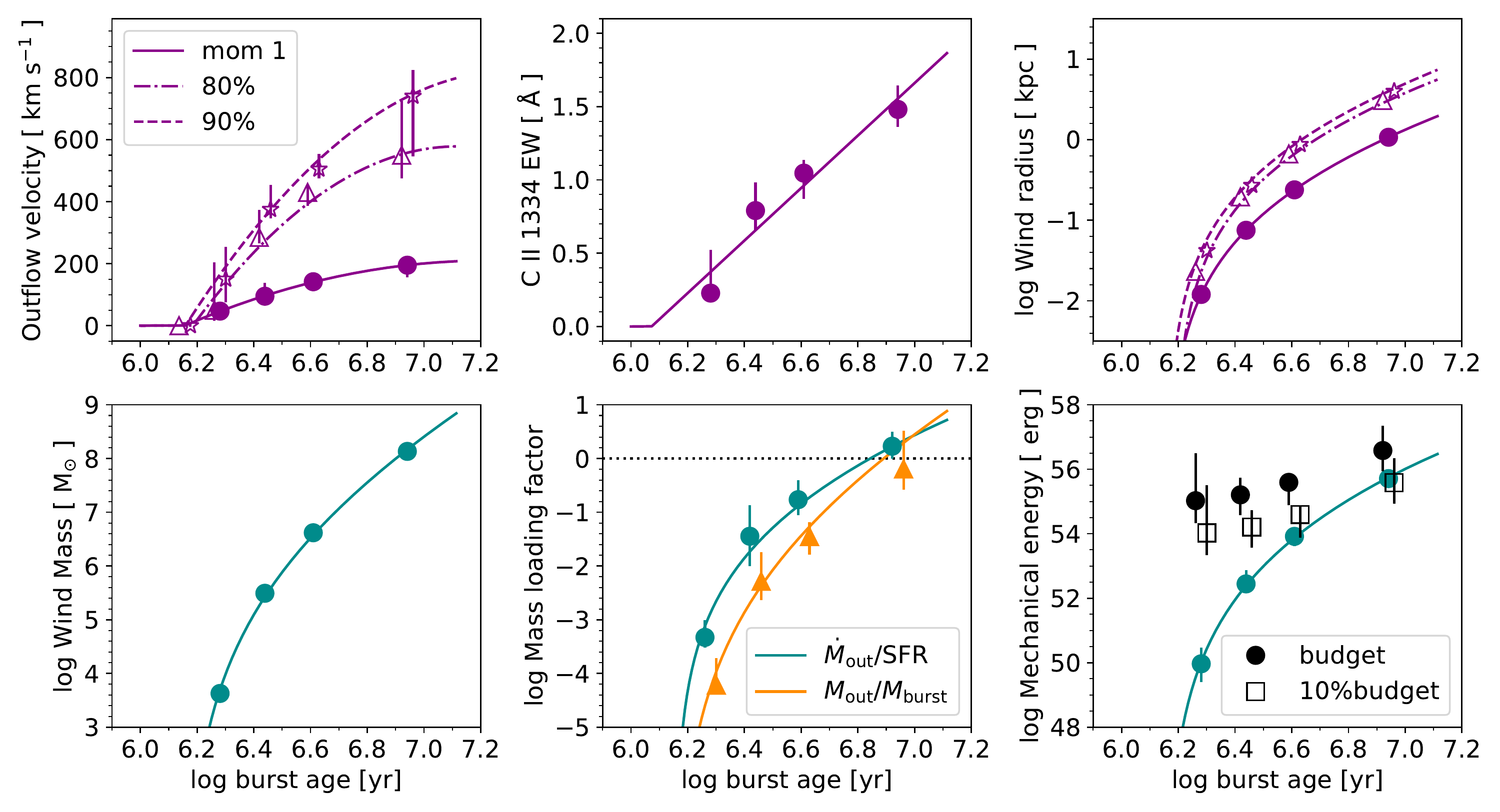}
    \caption{The evolution of the wind. Upper left shows the outflow velocity as probed by the first moment, and where the absorption crosses 80\% and 90\% of the continuum.  Each age corresponds to a stack in Figure~\ref{fig:abslines}. Lines show the best-fitting functions.  Upper centre shows the \ion{C}{ii} equivalent width, and a corresponding fit.   Upper right shows the wind radius, obtained by integrating the fits in the upper left panel with respect to time.  The lower left panel shows the mass in the wind.  The lower centre shows the mass-loading factor, defined as $\dot M_\mathrm{wind}/\mathrm{SFR}$, and the equivalent dimensionless property of $M_\mathrm{wind}$ divided by the stellar mass formed in the burst.  The dotted horizontal line shows the point at which $\eta$ surpasses unity.  The lower right panel shows the kinetic energy in the wind, assuming all gas is moving at the first moment of velocity.  The solid black markers show the integrated mechanical energy returned to the ISM by stellar winds and supernova, estimated from the stellar population synthesis modeling. Open squares show the same points scaled down to 10\%. Error-bars on the first two panels are purely observational, but on the remaining figures are propagated through the velocity-time integral derived in Section~\ref{sect:results}.  In some panels the data-points are offset in age by 0.01 dex for clarity, but solid circles always show the exact value.}  
    \label{fig:winds}
\end{figure*}

The main properties describing the evolution of the winds are shown in Figure~\ref{fig:winds}.  We begin by showing the outflow velocity in the upper left panel, for which we illustrate the evolution in the first moment, and velocities corresponding to 80\% and 90\% residual intensity. The lines are simply polynomial fits that allow us to integrate the curves, and do not contain any physics; this will be the subject of a followup paper.  The first moment of the velocity appears to start close to static, and accelerates to almost 200~\kms\ in 10~Myr. This acceleration is roughly constant with logarithmic time, implying an increasing acceleration of the wind. The outer edge of the detectable wind is accelerated faster, and the curves for the 80\% and 90\% absorption indicate that the {absorbing material} may be `stretching'.  This observationally-determined result is consistent with the theoretical picture of \citet{delaCruz.2021} who show that the low density material accelerates faster than the high density gas.

The upper central panel shows the evolution of the \ion{C}{ii} EW, which increases from $\approx 0.2$~\AA\ to 1.5~\AA\ over the same 10~Myr.  We fit a polynomial to these EWs, which intersects 0~\AA\ at almost the same age as the velocity ($\sim 10^{6.1}$~yr).

The upper right panel shows the evolution of $r_\mathrm{wind}$.  These curves are derived by integrating the velocity curves in the upper left figure, from a starting time of $\log(t/\mathrm{yr}) = 6.16$ (when $v_\mathrm{wind}=0$).  From here on, we work only with these fitting formulae, as we need higher time resolution than that provided by the four individual bins to evaluate the integrals. These curves show how the wind is blown outward.  At times of $\approx 1$~Myr the centroid of the absorption is consistent with the sizes of individual star clusters ($<10$pc), but grows rapidly to reach $\approx 1$~kpc after about 10~Myr. The faster moving edge of the wind of course moves further, reaching distances of $\approx 7$~kpc in the same time.  The radius of the wind can be described by the formula
\begin{equation} 
\log (r_\mathrm{wind}/\mathrm{kpc}) = \alpha x^\beta + \gamma  
\end{equation} 
where $x = \log(t)-6.16$ and $t$ is the burst age in years.  The radius described by the first moment of absorption can be calculated using $ \{ \alpha,\beta,\gamma \} = \{ 10.0, ~0.124,$ ~--9.69 \}\footnote{The 80\% and 90\% of the absorption can be calculated using $ \{ \alpha,\beta,\gamma \} = \{ 24.2, ~0.0494$, ~--23.4 \} and $ \{ 11.2, ~0.112$, ~10.3\}, respectively.}. We caution readers not to extend this function to times significantly greater than shown.

With an estimate of the radius we can proceed to make a number of further estimates for the properties of the wind.  Firstly we assume that the first moment describes the average velocity, and approximately traces the bulk of the mass. We then adopt the covering fraction of the wind (\fcov) to be unity.  It is not certain that this is the case, and \fcov\ may be lower on average, especially at higher velocities.  Indeed the absorption lines are very shallow at early times (Figure~\ref{fig:abslines}), probably because of incomplete covering.  However, at later times residual intensities fall below 50\%, suggesting the uncertainties that arise from \fcov~$<1$ are unlikely to exceed factors of 2. We estimate the column density of gas using the method described in Section~\ref{sect:windmodel}, and with the nebular metallicities from Figure~\ref{fig:genprops} we calculate the mass in hydrogen and helium, assuming solar abundance patterns \citep{Asplund.2009}.  

With the mass in the wind (shown for completeness in the lower left panel) and the dynamical timescale, we calculate the mass outflow rate: $\dot M_\mathrm{wind} \equiv M_\mathrm{wind}/t_\mathrm{dyn}$.  When normalized by the SFR (Figure~\ref{fig:genprops}) this gives the mass-loading factor ($\eta$), simply defined as the dimensionless ratio $\eta \equiv \dot M_\mathrm{wind}/\mathrm{SFR}$. We show the evolution of $\eta$ (now just using the first moment of velocity, with SFR calculated from \halpha) in the lower central panel.  We also calculate a similar dimensionless ratio of $M_\mathrm{wind}$ normalized by the mass formed in the starburst: this shows a very similar evolution to $\eta$ although runs somewhat steeper. As with other relations, $\eta$ is small at early times as little cool material is incorporated into the wind, but increases rapidly with time.  After $\approx 10$~Myr, $\eta$ exceeds unity, implying mass is removed from the star forming regions faster than it is converted into stars.  

With the mass in the wind and a characteristic velocity, we finally calculate its kinetic energy as $1/2 M_\mathrm{wind} v_\mathrm{wind}^2$, and show this in the lower right panel of Figure~\ref{fig:winds}. Again it appears similar to other figures but occupies a larger dynamic range because of the dependence on $v_\mathrm{wind}^2$.  Based upon the Starburst99 population modeling (Section~\ref{sect:model}) we estimate the recent star formation history (age of the starburst population) from which we compute the total mechanical energy deposited in the ISM by winds from massive stars and supernova explosions. We show this `energy budget' as the black solid circles in the lower right panel, which are substantially above the energy found in the wind at all times. The difference amounts to several orders of magnitude at the earliest times, but converges with time and is reduced to a factor of $\sim 10$ at 10~Myr.  We illustrate this simply by dividing each estimate by 10 in the open squares.  The large discrepancy at early times would be expected if there is a time delay in the coupling of stellar feedback and winds.  This could be because it takes time for ejecta to thermalize and expand, causing a delay in accelerating cool gas; alternatively this may reflect the time required for cool gas to condense from the hotter phase.  Indeed in the central regions and at early times, the bulk of the energy is expected to reside in the very hot phase with temperature almost $10^8$~K \citep{Strickland.2009}.  The range spanned by this mechanical energy ratio can be compared to the lower panel of Figure~8 in \citet{Schneider.2020}, which contrasts the energy contained within the cool gas with that of the hotter phases -- they find roughly 8\% of the available energy in the $T<2\times10^4$~K gas at radius of 4~kpc and time of 35~Myr.  While both estimates are subject to significant uncertainties, the similar values are encouraging. 

\section{Discussion}\label{sect:disc}

In this work we contrast the outflow velocity of galaxy winds with the evolutionary stage of the starburst, showing that winds appear to be accelerated over timescales of $\approx 10$~Myr.  Most studies of galactic outflows treat them a steady-state phenomena, while gas observed to be moving \emph{must have been accelerated}.  To our knowledge, this is the first time that the temporal evolution of winds has been explored.

\subsection{Implications and Comparisons}

We estimate that in spherically averaged galaxy winds, mass loading factors are driven to values close to 1 after times of $\approx 10$~Myr, when the winds have expanded to $r_\mathrm{wind}\approx 1$~kpc.  We further estimate that $\sim 10$~\% of the mechanical energy delivered by stellar feedback is to be found in the cool phase of the outflow after 10~Myr.  Analogous to the Big Bang, our data imply that the radius goes to zero at early times, and very small `launch radii' ($\ll 100$~pc) for galaxy winds have also been inferred \citep[e.g.][]{Chisholm.2018,Carr.2021} using completely different techniques and assumptions.  

The mass loading factors of $\eta  \approx 1-3$ and an escaping wind are quite comparable to those found using UV absorption \citep{Xu.2022classy}, and via emission lines such as \halpha\ \citep[][]{McQuinn.2019}.  In the more comparable methodology of Xu et al., the mass flow rate is calculated assuming $r_\mathrm{wind}$ scales with optical size of the starburst.  They use the results to investigate the dependence of $\eta$ on physical properties, finding an order-of-magnitude scatter.  One possible explanation for this dispersion is the evolutionary stage at which each galaxy wind is observed.  Indeed, one of the conclusions of our study is that the acceleration (or condensation) of the cool wind can be recovered from the data, and the wind properties estimated for a single galaxy must depend critically on when it is observed during the starburst episode.  This will be a function of the selection effects implicit in this study, and all others of similar nature. 

A particularly interesting find is that not only do winds accelerate with time but the EW of absorption lines also increases.  This EW is a direct result of the covering fraction and column density of absorbing metals, but in this sample it cannot be traced to an underlying effect of mass or metallicity (Figure~\ref{fig:genprops}).  Instead it appears that winds are becoming `thicker' on average, which could arise naturally for several reasons. As the wind develops, more cool material could be advected into the flow \citep[e.g.][]{Martin.2005,Gronke.2018}.  More evolved winds would also be larger and in a protracted star formation event, star clusters forming later would both dominate light and form behind a thicker envelope with covering fraction closer to 1.  \citet{Steidel.2010} also infer that if the outflows entrain dense cool clouds, then these must also expand as they move outwards, thereby increasing the covering fraction. 

\subsection{Caveats and points of improvement}

Naturally there are several caveats to this work that need to be addressed.  Most importantly the line-of-sight estimate of gas column density is oversimplified: we use the \ion{C}{ii}1334\AA\ absorption line to estimate column densities, assuming a unity covering fraction, in a stacking analysis, and adopt no ionization correction.  While the \ion{C}{ii} EW technique is calibrated against an optically thin gas tracer in similar galaxies, it is not expected to perform perfectly, and heavy saturation (especially at maximum absorption) may cause the total amount of gas to remain underestimated.  

There is no need to adopt the stacking analysis in principle, and we do so only because the archival COS spectra are not deep enough to do the experiment on individual galaxies.   Were the spectra better exposed, the method could also be improved by turning to the more numerous set of \ion{Si}{ii} absorption lines (1190 -- 1527~\AA) that have a range of oscillator strengths.  This ensemble of lines would allow us to estimate \fcov\ and column density, and therefore mass, in each velocity resolution element, as demonstrated in a similar strongly star-forming sample  by \citet{Rivera-Thorsen.2015}. These authors also showed that the maximum \fcov\ is very close to 1, and occurs at velocities where the bulk of the outflowing gas is piled up.  \fcov\ could still be overestimated at early times, which would artificially decrease the inferred gas masses in the first bin.  Given that the mass depends on the square of $r_\mathrm{wind}$, however, this is unlikely to affect the overall result because it enters at times where $r_\mathrm{wind}$ is smallest. 

The averaging process of stacking artificially `sphericalizes' galaxies and introduces an effective average covering term; in reality galaxy winds are often biconical may be modeled as such \citep[e.g.][]{Carr.2018}. However, it also remains to be demonstrated whether our galaxies, most of which are very compact, young, and intense starbursts, exhibit large-scale disks and biconical structures like, for example, M82.  Adopting the \ion{C}{ii}~$\lambda 1334$~\AA\ line alone means we sample only the cooler phase, and are insensitive to the warmer C$^{2+}$- and C$^{3+}$-bearing material. It is possible that the surprisingly small absorption in the earliest bin is due to this `over-ionization', and the cooler gas may indeed be condensing after a few Myr.  This may be tested with \ion{C}{iv} and \ion{Si}{iv} \cite[see e.g.][]{Xu.2022classy} doublets redwards in the UV, but our spectra are incomplete at this wavelength because of the redshift range of our sample. 

Finally, we note that \mstell\ and SFR are known to be correlated with outflow velocity \citep[e.g.][]{Heckman.2015}.  Trends of these properties with age are visible here (Figure~\ref{fig:genprops}), although we do not believe they span sufficient dynamic range to explain the wind evolution we see.  Nevertheless, the study would be improved with a large, homogeneously selected sample, for which deeper spectroscopy is available that would alleviate the need to stack data.  This is currently hard in even the best low-$z$ datasets, where small cleanly-selected samples ($N \lesssim 20$) do exist but are then subsumed (as here) into larger ones.

\subsection{Future applications}

One of the strengths of this method is that it does not rely upon morphological measurements, sizes, or inclinations, and is demonstrated here on unresolved data.  Thus if data could be obtained at sufficient depth it could be executed at high redshifts.  The physical size subtended by the COS aperture at $z=0.25$ (the average redshift of our sample) is $\approx 10$~kpc, which is equivalent to the coverage of a 1\farcs2 slit at $z=3$, and the aperture effects would be the same. Similar spectral modeling has already been performed in stacked spectra near $z\approx 3$ \citep{Steidel.2016, Cullen.2020}, and even in individual galaxies to $z$ of almost 5 \citep{Matthee.2022a.specmodel}, admittedly in extraordinarily deep data.  These methods can be applied anywhere the stellar continuum can be modeled and the absorption lines measured, although some high-$z$ samples (especially the more massive LBGs) may have more protracted star-formation histories, making it difficult to extract characteristic ages from the data.  At fainter magnitudes, we expect the same methods to easily be applied with spectrographs on Extremely Large Telescopes \citep[e.g. ELT/MOSAIC,][]{Sanchez-Janssen.2020}.  

We argue that these methods should be deployed in concert with existing techniques to most accurately derive the properties of galaxy winds.  When realized in large samples, the evolutionary scenarios described here may explain (some of) the variation of wind properties, and the observed dispersion on scaling relations.  Such updates would lead to more precise prescriptions of how much cool/warm gas is found in galaxy winds throughout star formation episodes. In turn, the observationally-derived constraints would provide foundational input for computer simulations that aim to describe the distribution of gas around galaxies, or that rely upon prescriptions for energy deposition while galaxies are experiencing their most rapid periods of growth.

\section*{Acknowledgements}

M.H. is Fellow of the Kunt \& Alice Wallenberg Foundation.  We thank Max Gronke, Axel Runnholm, and Claudia Scarlata for contributions to the project, and Angela Adamo, Antonello Calabr\`o, Mark Krumholz, Pierluigi Monaco and Mattia Sirressi for further useful discussions.  We are grateful to the anonymous referee for careful reading and thoughtful suggestions that have significantly improved the manuscript. 

\section*{Data Availability}

 All data are available in the Barbara A. Mikulski Archive for Space Telescopes (MAST) at  \href{https://archive.stsci.edu/hst/search.php}{https://archive.stsci.edu/hst/search.php}

\bsp	
\label{lastpage}

\begin{thebibliography}{}
\makeatletter
\relax
\def\mn@urlcharsother{\let\do\@makeother \do\$\do\&\do\#\do\^\do\_\do\%\do\~}
\def\mn@doi{\begingroup\mn@urlcharsother \@ifnextchar [ {\mn@doi@}
  {\mn@doi@[]}}
\def\mn@doi@[#1]#2{\def\@tempa{#1}\ifx\@tempa\@empty \href
  {http://dx.doi.org/#2} {doi:#2}\else \href {http://dx.doi.org/#2} {#1}\fi
  \endgroup}
\def\mn@eprint#1#2{\mn@eprint@#1:#2::\@nil}
\def\mn@eprint@arXiv#1{\href {http://arxiv.org/abs/#1} {{\tt arXiv:#1}}}
\def\mn@eprint@dblp#1{\href {http://dblp.uni-trier.de/rec/bibtex/#1.xml}
  {dblp:#1}}
\def\mn@eprint@#1:#2:#3:#4\@nil{\def\@tempa {#1}\def\@tempb {#2}\def\@tempc
  {#3}\ifx \@tempc \@empty \let \@tempc \@tempb \let \@tempb \@tempa \fi \ifx
  \@tempb \@empty \def\@tempb {arXiv}\fi \@ifundefined
  {mn@eprint@\@tempb}{\@tempb:\@tempc}{\expandafter \expandafter \csname
  mn@eprint@\@tempb\endcsname \expandafter{\@tempc}}}

\bibitem[\protect\citeauthoryear{{Abdurro'uf} et~al.,}{{Abdurro'uf}
  et~al.}{2022}]{SDSS.DR17}
{Abdurro'uf} et~al., 2022, \mn@doi [\apjs] {10.3847/1538-4365/ac4414}, \href
  {https://ui.adsabs.harvard.edu/abs/2022ApJS..259...35A} {259, 35}

\bibitem[\protect\citeauthoryear{{Asplund}, {Grevesse}, {Sauval}  \&
  {Scott}}{{Asplund} et~al.}{2009}]{Asplund.2009}
{Asplund} M.,  {Grevesse} N.,  {Sauval} A.~J.,   {Scott} P.,  2009, \mn@doi
  [\araa] {10.1146/annurev.astro.46.060407.145222}, \href
  {https://ui.adsabs.harvard.edu/abs/2009ARA&A..47..481A} {47, 481}

\bibitem[\protect\citeauthoryear{{Calabr{\`o}} et~al.,}{{Calabr{\`o}}
  et~al.}{2022}]{Calabro.2022}
{Calabr{\`o}} A.,  et~al., 2022, arXiv e-prints, \href
  {https://ui.adsabs.harvard.edu/abs/2022arXiv220614918C} {p. arXiv:2206.14918}

\bibitem[\protect\citeauthoryear{{Calzetti}, {Armus}, {Bohlin}, {Kinney},
  {Koornneef}  \& {Storchi-Bergmann}}{{Calzetti} et~al.}{2000}]{Calzetti.2000}
{Calzetti} D.,  {Armus} L.,  {Bohlin} R.~C.,  {Kinney} A.~L.,  {Koornneef} J.,
   {Storchi-Bergmann} T.,  2000, \mn@doi [\apj] {10.1086/308692}, \href
  {http://adsabs.harvard.edu/abs/2000ApJ...533..682C} {533, 682}

\bibitem[\protect\citeauthoryear{{Carr}, {Scarlata}, {Panagia}  \&
  {Henry}}{{Carr} et~al.}{2018}]{Carr.2018}
{Carr} C.,  {Scarlata} C.,  {Panagia} N.,   {Henry} A.,  2018, \mn@doi [\apj]
  {10.3847/1538-4357/aac48e}, \href
  {https://ui.adsabs.harvard.edu/abs/2018ApJ...860..143C} {860, 143}

\bibitem[\protect\citeauthoryear{{Carr}, {Scarlata}, {Henry}  \&
  {Panagia}}{{Carr} et~al.}{2021}]{Carr.2021}
{Carr} C.,  {Scarlata} C.,  {Henry} A.,   {Panagia} N.,  2021, \mn@doi [\apj]
  {10.3847/1538-4357/abc7c3}, \href
  {https://ui.adsabs.harvard.edu/abs/2021ApJ...906..104C} {906, 104}

\bibitem[\protect\citeauthoryear{{Chisholm}, {Tremonti}, {Leitherer}  \&
  {Chen}}{{Chisholm} et~al.}{2017}]{Chisholm.2017}
{Chisholm} J.,  {Tremonti} C.~A.,  {Leitherer} C.,   {Chen} Y.,  2017, \mn@doi
  [\mnras] {10.1093/mnras/stx1164}, \href
  {https://ui.adsabs.harvard.edu/abs/2017MNRAS.469.4831C} {469, 4831}

\bibitem[\protect\citeauthoryear{{Chisholm}, {Tremonti}  \&
  {Leitherer}}{{Chisholm} et~al.}{2018}]{Chisholm.2018}
{Chisholm} J.,  {Tremonti} C.,   {Leitherer} C.,  2018, \mn@doi [\mnras]
  {10.1093/mnras/sty2380}, \href
  {https://ui.adsabs.harvard.edu/abs/2018MNRAS.481.1690C} {481, 1690}

\bibitem[\protect\citeauthoryear{{Chisholm}, {Rigby}, {Bayliss}, {Berg},
  {Dahle}, {Gladders}  \& {Sharon}}{{Chisholm} et~al.}{2019}]{Chisholm.2019}
{Chisholm} J.,  {Rigby} J.~R.,  {Bayliss} M.,  {Berg} D.~A.,  {Dahle} H.,
  {Gladders} M.,   {Sharon} K.,  2019, \mn@doi [\apj]
  {10.3847/1538-4357/ab3104}, \href
  {https://ui.adsabs.harvard.edu/abs/2019ApJ...882..182C} {882, 182}

\bibitem[\protect\citeauthoryear{{Cullen} et~al.,}{{Cullen}
  et~al.}{2020}]{Cullen.2020}
{Cullen} F.,  et~al., 2020, \mn@doi [\mnras] {10.1093/mnras/staa1260}, \href
  {https://ui.adsabs.harvard.edu/abs/2020MNRAS.495.1501C} {495, 1501}

\bibitem[\protect\citeauthoryear{{Green} et~al.,}{{Green}
  et~al.}{2012}]{Green.2012}
{Green} J.~C.,  et~al., 2012, \mn@doi [\apj] {10.1088/0004-637X/744/1/60},
  \href {https://ui.adsabs.harvard.edu/abs/2012ApJ...744...60G} {744, 60}

\bibitem[\protect\citeauthoryear{{Gronke} \& {Oh}}{{Gronke} \&
  {Oh}}{2018}]{Gronke.2018}
{Gronke} M.,  {Oh} S.~P.,  2018, \mn@doi [\mnras] {10.1093/mnrasl/sly131},
  \href {https://ui.adsabs.harvard.edu/abs/2018MNRAS.480L.111G} {480, L111}

\bibitem[\protect\citeauthoryear{{Hayes}, {Runnholm}, {Gronke}  \&
  {Scarlata}}{{Hayes} et~al.}{2021}]{Hayes.2021}
{Hayes} M.~J.,  {Runnholm} A.,  {Gronke} M.,   {Scarlata} C.,  2021, \mn@doi
  [\apj] {10.3847/1538-4357/abd246}, \href
  {https://ui.adsabs.harvard.edu/abs/2021ApJ...908...36H} {908, 36}

\bibitem[\protect\citeauthoryear{{Heckman} et~al.,}{{Heckman}
  et~al.}{2011}]{Heckman.2011}
{Heckman} T.~M.,  et~al., 2011, \mn@doi [\apj] {10.1088/0004-637X/730/1/5},
  \href {http://adsabs.harvard.edu/abs/2011ApJ...730....5H} {730, 5}

\bibitem[\protect\citeauthoryear{{Heckman}, {Alexandroff}, {Borthakur},
  {Overzier}  \& {Leitherer}}{{Heckman} et~al.}{2015}]{Heckman.2015}
{Heckman} T.~M.,  {Alexandroff} R.~M.,  {Borthakur} S.,  {Overzier} R.,
  {Leitherer} C.,  2015, \mn@doi [\apj] {10.1088/0004-637X/809/2/147}, \href
  {http://adsabs.harvard.edu/abs/2015ApJ...809..147H} {809, 147}

\bibitem[\protect\citeauthoryear{{Henry}, {Scarlata}, {Martin}  \&
  {Erb}}{{Henry} et~al.}{2015}]{Henry.2015}
{Henry} A.,  {Scarlata} C.,  {Martin} C.~L.,   {Erb} D.,  2015, \mn@doi [\apj]
  {10.1088/0004-637X/809/1/19}, \href
  {http://adsabs.harvard.edu/abs/2015ApJ...809...19H} {809, 19}

\bibitem[\protect\citeauthoryear{{Izotov}, {Worseck}, {Schaerer}, {Guseva},
  {Thuan}, {Fricke}  \& {Orlitov{\'a}}}{{Izotov} et~al.}{2018}]{Izotov.2018}
{Izotov} Y.~I.,  {Worseck} G.,  {Schaerer} D.,  {Guseva} N.~G.,  {Thuan} T.~X.,
   {Fricke} Verhamme A.,   {Orlitov{\'a}} I.,  2018, \mn@doi [\mnras]
  {10.1093/mnras/sty1378}, \href
  {https://ui.adsabs.harvard.edu/abs/2018MNRAS.478.4851I} {478, 4851}

\bibitem[\protect\citeauthoryear{{James}, {Aloisi}, {Heckman}, {Sohn}  \&
  {Wolfe}}{{James} et~al.}{2014}]{James.2014}
{James} B.~L.,  {Aloisi} A.,  {Heckman} T.,  {Sohn} S.~T.,   {Wolfe} M.~A.,
  2014, \mn@doi [\apj] {10.1088/0004-637X/795/2/109}, \href
  {http://adsabs.harvard.edu/abs/2014ApJ...795..109J} {795, 109}

\bibitem[\protect\citeauthoryear{{Kennicutt} \& {Evans}}{{Kennicutt} \&
  {Evans}}{2012}]{Kennicutt.2012}
{Kennicutt} R.~C.,  {Evans} N.~J.,  2012, \mn@doi [\araa]
  {10.1146/annurev-astro-081811-125610}, \href
  {https://ui.adsabs.harvard.edu/abs/2012ARA&A..50..531K} {50, 531}

\bibitem[\protect\citeauthoryear{{Leitherer} et~al.,}{{Leitherer}
  et~al.}{1999}]{Leitherer.1999}
{Leitherer} C.,  et~al., 1999, \mn@doi [\apjs] {10.1086/313233}, \href
  {http://adsabs.harvard.edu/abs/1999ApJS..123....3L} {123, 3}

\bibitem[\protect\citeauthoryear{{Leitherer}, {Ekstr{\"o}m}, {Meynet},
  {Schaerer}, {Agienko}  \& {Levesque}}{{Leitherer}
  et~al.}{2014}]{Leitherer.2014}
{Leitherer} C.,  {Ekstr{\"o}m} S.,  {Meynet} G.,  {Schaerer} D.,  {Agienko}
  K.~B.,   {Levesque} E.~M.,  2014, \mn@doi [\apjs]
  {10.1088/0067-0049/212/1/14}, \href
  {http://adsabs.harvard.edu/abs/2014ApJS..212...14L} {212, 14}

\bibitem[\protect\citeauthoryear{{Marino} et~al.,}{{Marino}
  et~al.}{2013}]{Marino.2013}
{Marino} R.~A.,  et~al., 2013, \mn@doi [\aap] {10.1051/0004-6361/201321956},
  \href {https://ui.adsabs.harvard.edu/abs/2013A&A...559A.114M} {559, A114}

\bibitem[\protect\citeauthoryear{{Martin}}{{Martin}}{2005}]{Martin.2005}
{Martin} C.~L.,  2005, \mn@doi [\apj] {10.1086/427277}, \href
  {http://adsabs.harvard.edu/abs/2005ApJ...621..227M} {621, 227}

\bibitem[\protect\citeauthoryear{{Matthee} et~al.,}{{Matthee}
  et~al.}{2022}]{Matthee.2022a.specmodel}
{Matthee} J.,  et~al., 2022, \mn@doi [\aap] {10.1051/0004-6361/202142187},
  \href {https://ui.adsabs.harvard.edu/abs/2022A&A...660A..10M} {660, A10}

\bibitem[\protect\citeauthoryear{{McQuinn}, {van Zee}  \& {Skillman}}{{McQuinn}
  et~al.}{2019}]{McQuinn.2019}
{McQuinn} K. B.~W.,  {van Zee} L.,   {Skillman} E.~D.,  2019, \mn@doi [\apj]
  {10.3847/1538-4357/ab4c37}, \href
  {https://ui.adsabs.harvard.edu/abs/2019ApJ...886...74M} {886, 74}

\bibitem[\protect\citeauthoryear{{Rivera-Thorsen} et~al.,}{{Rivera-Thorsen}
  et~al.}{2015}]{Rivera-Thorsen.2015}
{Rivera-Thorsen} T.~E.,  et~al., 2015, \mn@doi [\apj]
  {10.1088/0004-637X/805/1/14}, \href
  {http://adsabs.harvard.edu/abs/2015ApJ...805...14R} {805, 14}

\bibitem[\protect\citeauthoryear{{Salpeter}}{{Salpeter}}{1955}]{Salpeter.1955}
{Salpeter} E.~E.,  1955, \mn@doi [\apj] {10.1086/145971}, \href
  {https://ui.adsabs.harvard.edu/abs/1955ApJ...121..161S} {121, 161}

\bibitem[\protect\citeauthoryear{{S{\'a}nchez-Janssen}
  et~al.,}{{S{\'a}nchez-Janssen} et~al.}{2020}]{Sanchez-Janssen.2020}
{S{\'a}nchez-Janssen} R.,  et~al., 2020, in Society of Photo-Optical
  Instrumentation Engineers (SPIE) Conference Series. p. 1144725 (\mn@eprint
  {arXiv} {2012.08393}), \mn@doi{10.1117/12.2561222}

\bibitem[\protect\citeauthoryear{{Schneider}, {Ostriker}, {Robertson}  \&
  {Thompson}}{{Schneider} et~al.}{2020}]{Schneider.2020}
{Schneider} E.~E.,  {Ostriker} E.~C.,  {Robertson} B.~E.,   {Thompson} T.~A.,
  2020, \mn@doi [\apj] {10.3847/1538-4357/ab8ae8}, \href
  {https://ui.adsabs.harvard.edu/abs/2020ApJ...895...43S} {895, 43}

\bibitem[\protect\citeauthoryear{{Sirressi} et~al.,}{{Sirressi}
  et~al.}{2022}]{Sirressi.2022}
{Sirressi} M.,  et~al., 2022, \mn@doi [\mnras] {10.1093/mnras/stab3774}, \href
  {https://ui.adsabs.harvard.edu/abs/2022MNRAS.510.4819S} {510, 4819}

\bibitem[\protect\citeauthoryear{{Somerville} \& {Dav{\'e}}}{{Somerville} \&
  {Dav{\'e}}}{2015}]{Somerville2015}
{Somerville} R.~S.,  {Dav{\'e}} R.,  2015, \mn@doi [\araa]
  {10.1146/annurev-astro-082812-140951}, \href
  {http://adsabs.harvard.edu/abs/2015ARA%26A..53...51S} {53, 51}

\bibitem[\protect\citeauthoryear{{Steidel}, {Erb}, {Shapley}, {Pettini},
  {Reddy}, {Bogosavljevi{\'c}}, {Rudie}  \& {Rakic}}{{Steidel}
  et~al.}{2010}]{Steidel.2010}
{Steidel} C.~C.,  {Erb} D.~K.,  {Shapley} A.~E.,  {Pettini} M.,  {Reddy} N.,
  {Bogosavljevi{\'c}} M.,  {Rudie} G.~C.,   {Rakic} O.,  2010, \mn@doi [\apj]
  {10.1088/0004-637X/717/1/289}, \href
  {http://adsabs.harvard.edu/abs/2010ApJ...717..289S} {717, 289}

\bibitem[\protect\citeauthoryear{{Steidel}, {Strom}, {Pettini}, {Rudie},
  {Reddy}  \& {Trainor}}{{Steidel} et~al.}{2016}]{Steidel.2016}
{Steidel} C.~C.,  {Strom} A.~L.,  {Pettini} M.,  {Rudie} G.~C.,  {Reddy} N.~A.,
    {Trainor} R.~F.,  2016, \mn@doi [\apj] {10.3847/0004-637X/826/2/159}, \href
  {http://adsabs.harvard.edu/abs/2016ApJ...826..159S} {826, 159}

\bibitem[\protect\citeauthoryear{{Strickland} \& {Heckman}}{{Strickland} \&
  {Heckman}}{2009}]{Strickland.2009}
{Strickland} D.~K.,  {Heckman} T.~M.,  2009, \mn@doi [\apj]
  {10.1088/0004-637X/697/2/2030}, \href
  {http://adsabs.harvard.edu/abs/2009ApJ...697.2030S} {697, 2030}

\bibitem[\protect\citeauthoryear{{Worthey}}{{Worthey}}{1994}]{Worthey.1994}
{Worthey} G.,  1994, \mn@doi [\apjs] {10.1086/192096}, \href
  {https://ui.adsabs.harvard.edu/abs/1994ApJS...95..107W} {95, 107}

\bibitem[\protect\citeauthoryear{{Xu} et~al.,}{{Xu}
  et~al.}{2022}]{Xu.2022classy}
{Xu} X.,  et~al., 2022, \mn@doi [\apj] {10.3847/1538-4357/ac6d56}, \href
  {https://ui.adsabs.harvard.edu/abs/2022ApJ...933..222X} {933, 222}

\bibitem[\protect\citeauthoryear{{de la Cruz}, {Schneider}  \& {Ostriker}}{{de
  la Cruz} et~al.}{2021}]{delaCruz.2021}
{de la Cruz} L.~M.,  {Schneider} E.~E.,   {Ostriker} E.~C.,  2021, \mn@doi
  [\apj] {10.3847/1538-4357/ac04ac}, \href
  {https://ui.adsabs.harvard.edu/abs/2021ApJ...919..112D} {919, 112}

\makeatother
\end{thebibliography}
\end{document}